\documentclass[preprint,prd,tightenlines,nofootinbib,floatfix,superscriptaddress]{revtex4-1}
\usepackage{color}
\usepackage{graphicx}
\definecolor{red}{rgb}{1,0,0}
\def\lesssim{\ \hbox{\raise 2pt \hbox{$<$} \kern -13pt
                     \lower 3pt \hbox{$\sim$}}\ }
\def\greatersim{\ \hbox{\raise 2pt \hbox{$>$} \kern -13pt
                     \lower 3pt \hbox{$\sim$}}\ }

\def\pythia{{\sc Pythia}}

\input epsf.tex
\def\desepsf(#1 width #2){\epsfxsize=#2 \epsfbox{#1}}

\usepackage{amsmath,bm}

\begin{document}
\hspace*{13.2 cm} {\small DESY 15-147} \\
\hspace*{13.0 cm} {\small RAL-P-2015-007}
\vspace*{1.4 cm} 
\title{Treating jet correlations in 
high pile-up    at hadron   colliders}
\author{F.\ Hautmann} 
\affiliation{Rutherford Appleton Laboratory $\&$  
University of   Southampton}
\affiliation{Theoretical Physics, 
University of Oxford}  
\author{H.\ Jung}
\affiliation{Deutsches Elektronen Synchrotron}
\affiliation{Elementaire Deeltjes Fysica, Universiteit Antwerpen}
\author{H.\ Van Haevermaet}
\affiliation{Elementaire Deeltjes Fysica, Universiteit Antwerpen}
\begin{abstract}
Experiments in  the 
high-luminosity runs at the 
Large Hadron Collider    
face the   
challenges of very large pile-up. Primary techniques to deal with 
this  are based on precise 
vertex and track reconstruction. Outside tracker  acceptances,
however, lie regions of   interest for  many aspects of the
LHC physics program. 
We explore 
complementary  
    approaches to pile-up treatment     
and propose  a data-driven   jet-mixing   
method     
which can be used outside 
tracker acceptances without  
depending 
 on Monte Carlo generators. 
The method can be applied to treat 
correlation observables and  take into 
account, besides the jet transverse 
momentum pedestal, effects  of   
 hard jets from pile-up. 
\end{abstract} 


\maketitle

{\em 1.~Introduction}.  
Experiments at hadron colliders operating with  
very high luminosity face the challenge  of  
pile-up, namely, a very  large number   of 
overlaid 
hadron-hadron collisions per bunch crossing. 
At the Large Hadron Collider (LHC), in  Run I data  
the pile-up is about 20 $ p p $ collisions on average,  
while it  reaches  the level of 
over 50 at Run II, and increases  
 for  higher-luminosity 
runs~\cite{ATLAS:2014cva,ATLAS:2014nea,TheATLAScollaboration:2013pia,Marshall:2014mza,CMS:2014ata,CMS:2013wea,Ghosh:2015raa,snowmass2013,Haas:2014talk,Hildreth:2014talk,Fartoukh:2014nga}. 
In regions covered by tracking detectors, 
advanced vertexing techniques have been developed 
to deal with environments characterized by 
high pile-up.  More generally, experiments rely on 
Monte Carlo simulations which include pile-up 
for comparisons  with  data.  However, 
this  introduces a significant model 
dependence,  especially in regions where 
no detailed and precise measurements are available 
to constrain Monte Carlo modeling. 

In this paper we propose a different approach 
to treating  high pile-up, with a view to  employing 
data-driven methods  rather than Monte Carlo methods.   
Our main focus is to deal with  potentially large probabilities 
that jets with high transverse momenta  are produced from 
pile-up events independent of the primary interaction vertex, in  a region 
where tracking devices are not available to identify pile-up jets. 
A typical application would be  Higgs production by vector boson 
fusion, where the associated   jets may be  produced  
  outside the tracking detector acceptances. 
The issue we address is thus quite different from the issues   that 
 most of the existing methods 
for pile-up treatment are designed 
to  deal with, which are 
 the jet transverse momentum pedestal,  due to the 
 bias in the jet transverse 
momentum from  added pile-up 
particles in the jet cone, and the 
clustering  into jets of  overlapping 
soft particles from pile-up. 

In what follows 
we will therefore use standard  existing  methods to treat soft particles 
and the jet pedestal, 
and devise new approaches to 
 tackle  the issue of misidentification  which arises, in addition,  
  in cases   where precise  tracking  and vertexing 
are   not feasible. 
 The aim is to look for  methods which treat 
pile-up without spoiling the 
physics of the signal process and which can be used 
outside the tracking detector acceptances without 
depending on Monte Carlo modeling.  To this end, we 
suggest using minimum bias (or jet) samples recorded  
from  data  in high pile-up runs and applying 
event-mixing techniques to relate, via these data samples,  
the ``true" signal to the signal measured in  high pile-up.

The approach does not address the question of a full 
detector simulation including pile-up. Rather, it focuses 
on how to extract physics signals with the least 
dependence on pile-up simulation, and how to use real 
data, rather than Monte Carlo events, at physics object 
level. 

The proposed method applies to the regime of high pile-up 
which is 
relevant for the LHC   as well as 
 for future high-luminosity colliders. 
It is designed to treat not only inclusive variables but also 
correlations.  One of the  features of the method is that 
it does not require data-taking in dedicated runs at  low 
pile-up. Rather, the data required for event mixing are 
recorded at the same time as the signal events in high 
pile-up runs, so that there is no loss in luminosity. 

We will  illustrate the approach using   
Drell-Yan lepton pair production associated with jets 
 as a case study. We discuss two main physical consequences 
of pile-up collisions, the  bias in the jet 
transverse momentum due to pile-up particles in the jet cone, 
and the misidentification of high transverse momentum jets 
from independent pile-up events. 
The method is general and can straightforwardly be extended 
to a large variety of processes affected by pile-up. 

\begin{figure}[htb]
\includegraphics[scale=0.5]{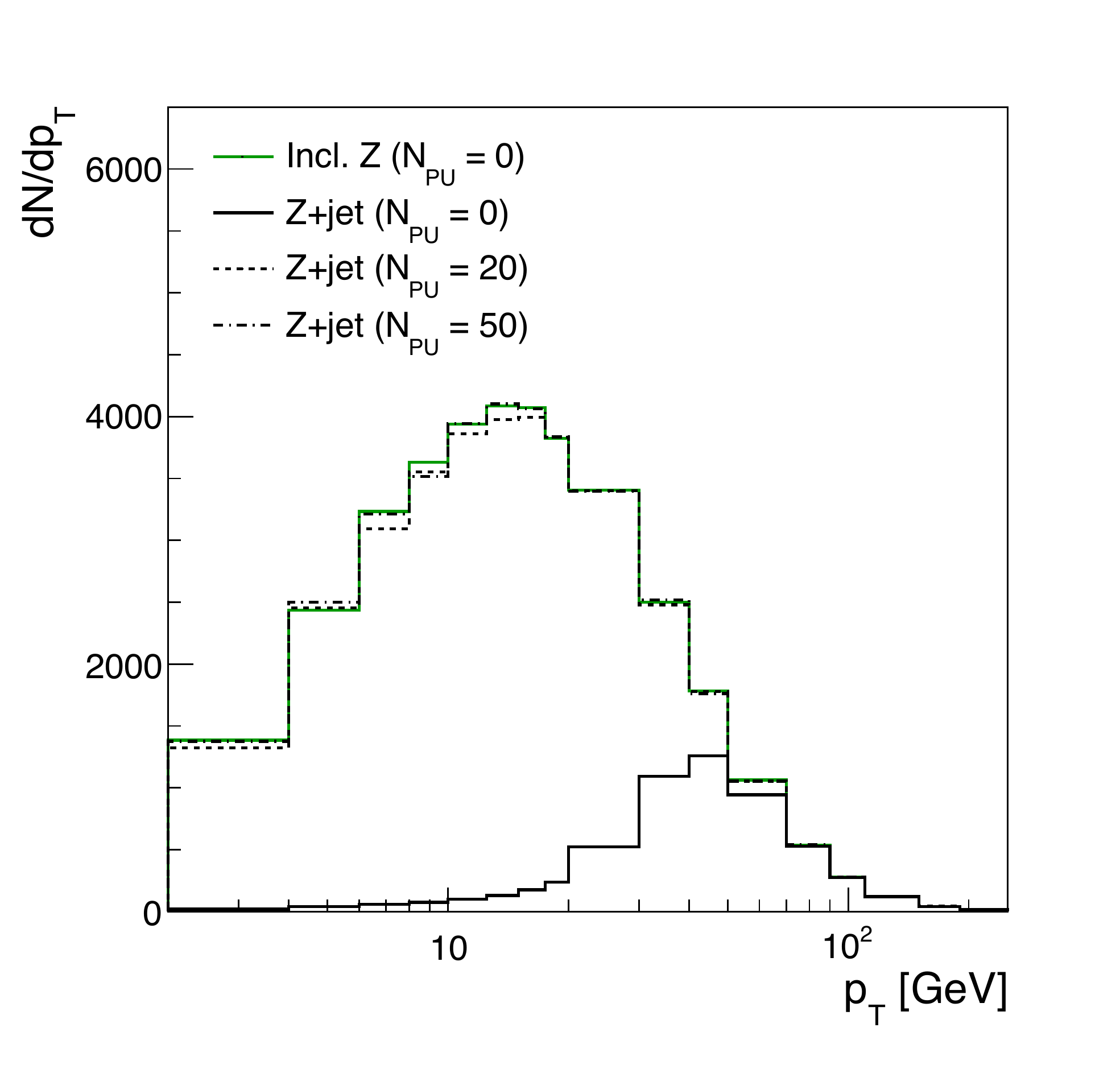}
  \caption{\it 
Effect of pile-up on 
the  $Z$-boson 
 transverse momentum $p_T$ in 
$Z$-boson + jet production at the LHC.}
\label{fig:plp_fig1}
\end{figure}

{\em 2.~Drell-Yan plus jets at high pile-up as a case 
study}.  
Let us consider the associated 
production of  a 
Drell-Yan lepton pair  
via  $Z$-boson exchange and  a  jet. 
We take the jet transverse momentum 
and rapidity to be 
$ p_T^{({\rm{jet}})} > 30 $ GeV, 
$ | \eta^{({\rm{jet}})}  |  <  4.5 $, 
and the boson invariant mass 
and rapidity  to be 
 60  GeV $  < m^{({\rm{boson}})} < $ 120 GeV, 
 $ | \eta^{({\rm{boson}})}  |  <  2 $. 
Event samples are generated 
by \pythia\ 8~\cite{Sjostrand:2007gs}  
with the 4C tune~\cite{Corke:2010yf}  
for the different  scenarios of zero  pile-up and 
 $N_{\rm{PU}}$ additional  $pp$  collisions at $\sqrt{s} = 13$ TeV.  
We reconstruct jets with the 
anti-$k_T$ algorithm~\cite{Cacciari:2008gp} with 
distance parameter $R = 0.5$. 
Results for the 
spectrum in the 
transverse momentum  
$p_T$  of the $Z$-boson, for $Z$ + jet events,     are shown in  
Fig.~\ref{fig:plp_fig1} for $N_{\rm{PU}} = 0$, 
$N_{\rm{PU}} = 20$ and $N_{\rm{PU}} = 50$.  For comparison we also 
show the inclusive $N_{\rm{PU}} = 0$ 
$Z$-boson spectrum. 

We see   from  
Fig.~\ref{fig:plp_fig1} 
 that the effects of pile-up 
on 
$Z$-boson  + jet scenarios 
are  large. As a result of pile-up 
the shape  of the 
$p_T$  spectrum is  
changed and the peak is 
shifted  to lower values. This can 
be interpreted by noting that,  
as the $Z$ + jet event sample 
 becomes dominated by 
pile-up collisions, 
even with 
the jet transverse momentum 
selection cut 
$ p_T^{({\rm{jet}})} > 30 $ GeV   
  the 
 $Z$-boson $p_T$ distribution 
in boson + jet  events will 
tend to 
approach the inclusive  Drell-Yan  
spectrum, given by the 
solid green curve.

More precisely, we  can identify two main 
implications of pile-up collisions: 
a large bias in the jet transverse 
momentum due to added pile-up 
particles in the jet cone leading to a jet pedestal,  and 
a large probability  that jets with 
high transverse momentum come 
from independent pile-up events.

\begin{figure}[htb]
\includegraphics[scale=0.4]{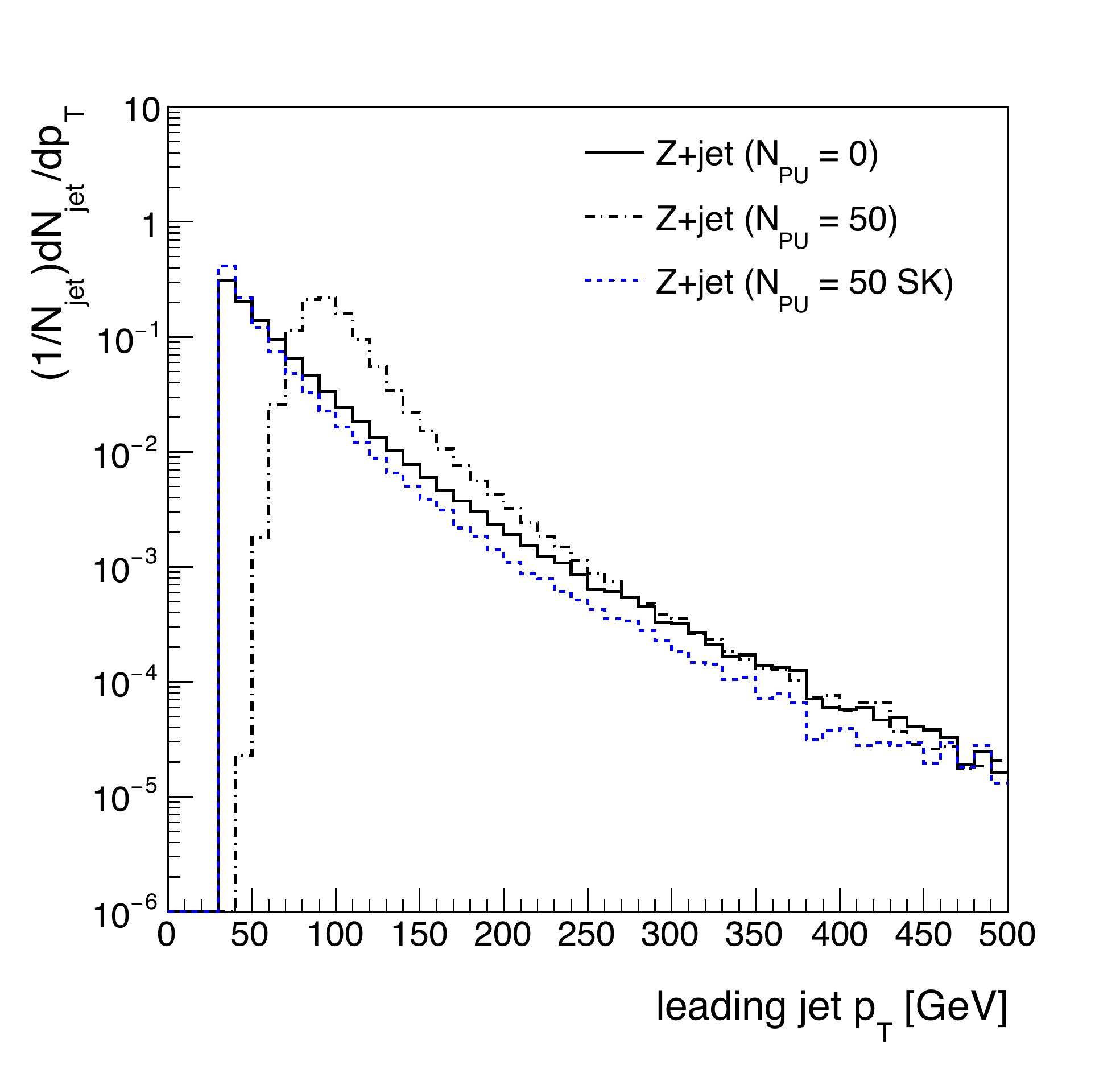}
\includegraphics[scale=0.4, trim=0cm 5.9cm 0cm 0cm, clip=true]{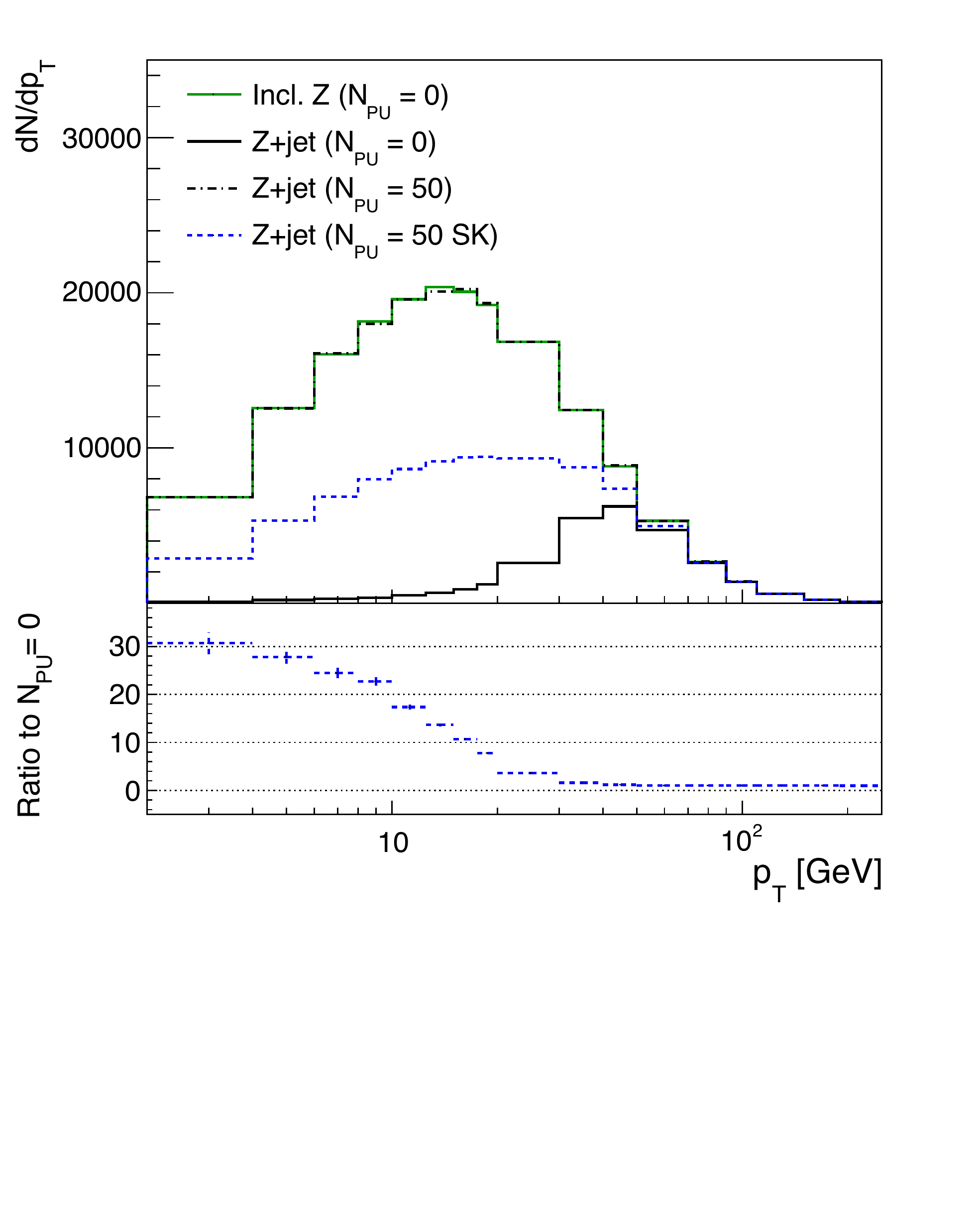}
  \caption{\it Application of SoftKiller~\protect\cite{Cacciari:2014gra}    
to $Z$-boson      +  jet production.  
Left: (a) the leading  jet $p_T$ spectrum; 
right: (b) the  $Z$-boson $p_T$ spectrum.}
\label{fig:plp_fig2}
\end{figure}

Several methods exist to deal 
with the jet $p_T$ pedestal. These include   
techniques based on the jet vertex 
fraction~\cite{TheATLAScollaboration:2013pia} 
and charged hadron  subtraction~\cite{CMS:2014ata,Kirschenmann:2014dla},  
the {\sc Puppi} method~\cite{Bertolini:2014bba}, 
the SoftKiller method~\cite{Cacciari:2014gra}. 
These methods correct for transverse 
momenta of individual particles,  but 
not for any mistagging. 
So do approaches inspired by 
jet substructure studies, such 
as jet cleansing~\cite{Krohn:2013lba}.  
In Fig.~\ref{fig:plp_fig2} 
we apply SoftKiller~\cite{Cacciari:2014gra}, 
a new event-wide particle-level pile-up removal method,
which can also be 
used with calorimeter  information 
only. We present results at zero pile-up ($N_{\rm{PU}} = 0$), 
at  pile-up $N_{\rm{PU}} = 50$,  and the result at  
pile-up $N_{\rm{PU}} = 50$   with SoftKiller subtraction ($N_{\rm{PU}} = 50$ SK). 

Fig.~\ref{fig:plp_fig2}  illustrates 
different physical effects of pile-up 
in the leading 
jet  spectrum and in the 
$Z$-boson 
 spectrum. 
In Fig.~\ref{fig:plp_fig2}a  we 
compute the leading 
jet $p_T$ spectrum, and verify 
that  SoftKiller efficiently 
removes the jet 
pedestal  from pile-up:  
the zero pile-up jet spectrum 
(solid black curve) 
is shifted toward larger 
$p_T$ by pile-up collisions 
(dot-dashed black curve for 
$N_{\rm{PU}} = 50$) but 
the application of SoftKiller  
(dashed blue curve 
$N_{\rm{PU}} = 50$ SK)
corrects for this and 
restores  the original 
signal with very good approximation. 
In Fig.~\ref{fig:plp_fig2}b, on the 
other hand, we   
compute the  $Z$-boson 
$p_T$ spectrum. The solid black curve 
is the zero pile-up result, the dot-dashed black 
curve is the $N_{\rm{PU}} = 50$ result, and 
the dashed blue curve is the result of 
  applying 
SoftKiller.  In the  higher  $p_T$ 
part of the spectrum 
we observe that there is 
no  need for any correction. 
In contrast, in the  lower  
  $p_T$ part 
significant contributions are 
present 
from misidentified pile-up jets.  
These  are not corrected for, and 
need to be properly treated, particularly in regions outside tracker acceptances where 
vertexing techniques cannot be relied on to identify pile-up 
jets~\cite{CMS:2013wea}. We 
address this point  next.

{\em 3.~Uncorrelated event samples and jet mixing}.   
To treat effects beyond soft particles  and the 
jet  $p_T$ pedestal,  we employ 
event mixing 
techniques~\cite{Drijard:1984pe,Schael:2004ux,Schettler:2013oma,Dutta:2011gs} using 
uncorrelated  samples. 
The main idea is 
that the signal in the pile-up 
scenario is obtained 
 via  mixing 
from the 
signal without pile-up and 
a minimum bias sample of 
 data at high pile-up. 
Thus,  
to  identify the contribution of the 
high $p_T$ jets coming 
from independent pile-up events,  
we construct a signal plus pile-up 
 scenario in a data-driven manner. 
We do this 
by adding physics objects from
pileup background to event samples before selection criteria are applied.
The approach is designed to treat the 
region of high number $N_{\rm{PU}}$ 
 of pile-up events, 
where $(N_{\rm{PU}} + 1)  /  N_{\rm{PU}} 
  \approx  1$.  

\begin{figure}[htb]
\includegraphics[scale=0.4, trim=0cm 5.9cm 0cm 0cm, clip=true]{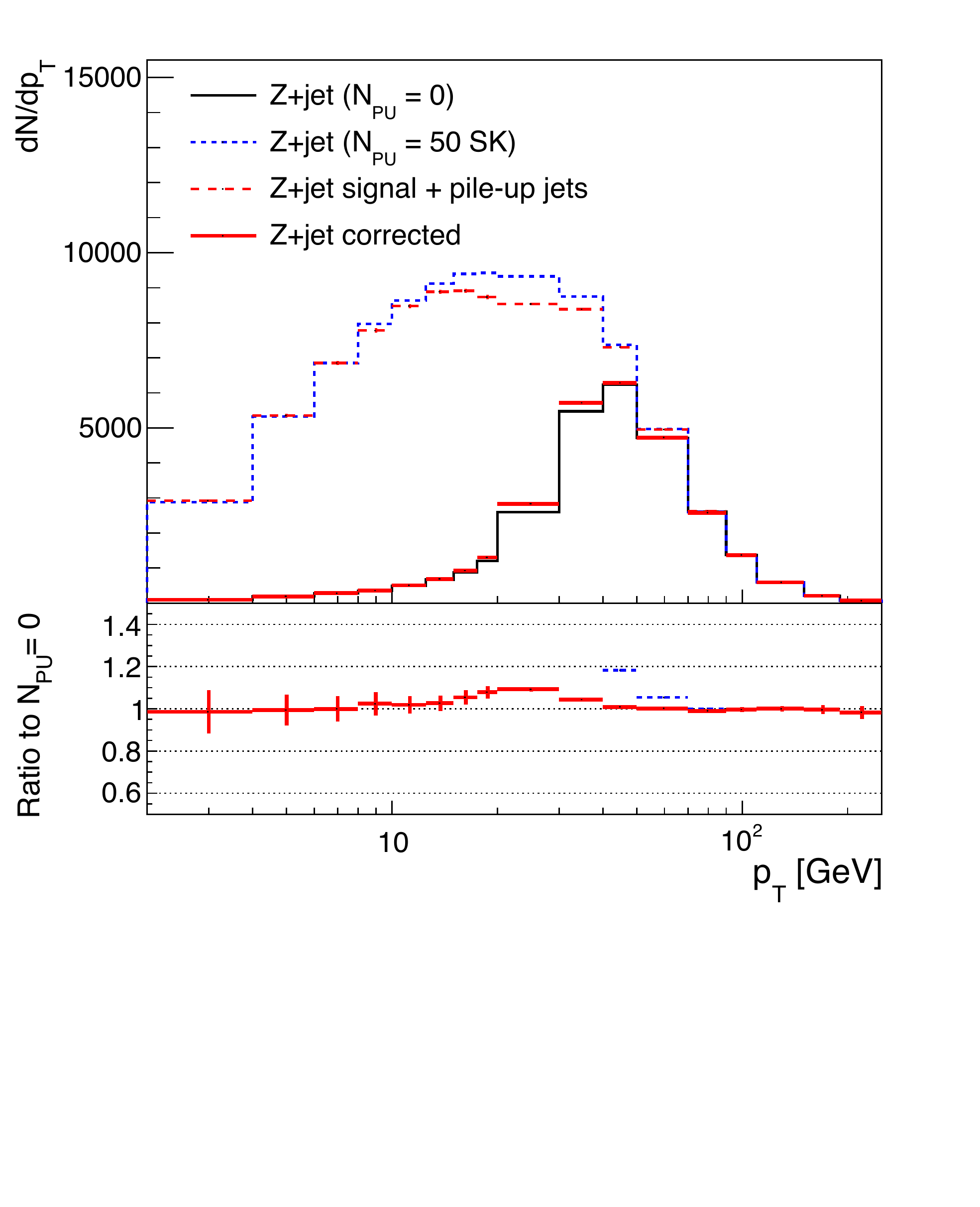}
\includegraphics[scale=0.4, trim=0cm 5.9cm 0cm 0cm, clip=true]{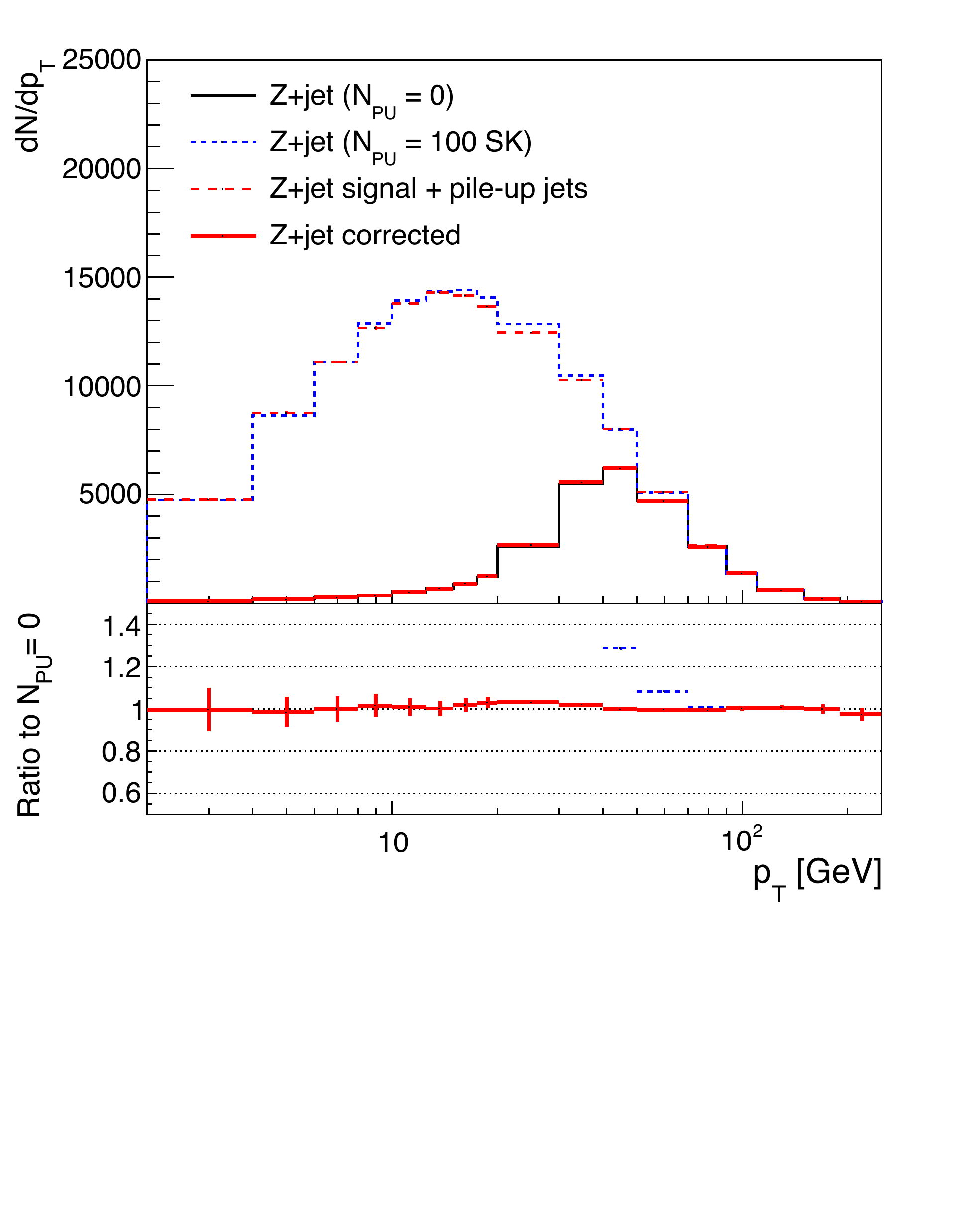}
  \caption{\it The  $Z$-boson $p_T$ spectrum  in $Z$ + jet production 
from the jet mixing method.  Left: (a) 
 $N_{\rm{PU}} = 50$;  right: (b) $N_{\rm{PU}} = 100$. 
}
\label{fig:plp_fig3}
\end{figure}

We illustrate the method by 
taking a sample  containing $N_{\rm{PU}}$ 
minimum bias 
events (which could be recorded data but we just take for illustration as 
Monte Carlo events), 
mixing this with 
the signal at zero  pile-up,  
and 
then 
 requiring  a jet   
with $ p_T^{({\rm{jet}})} > 30 $ GeV, 
$ | \eta^{({\rm{jet}})}  |  <  4.5 $. 
We extract  the unbiased signal 
 without relying 
on Monte Carlo  algorithms.

Fig.~\ref{fig:plp_fig3} reports the 
result of carrying 
out  this procedure,   
 for 
$N_{\rm{PU}} = 50$ and $N_{\rm{PU}} = 100$. 
Here 
the solid black 
curve is the ``true" 
$Z$-boson plus jet signal.  
\begin{figure}[htb]
\includegraphics[scale=0.6]{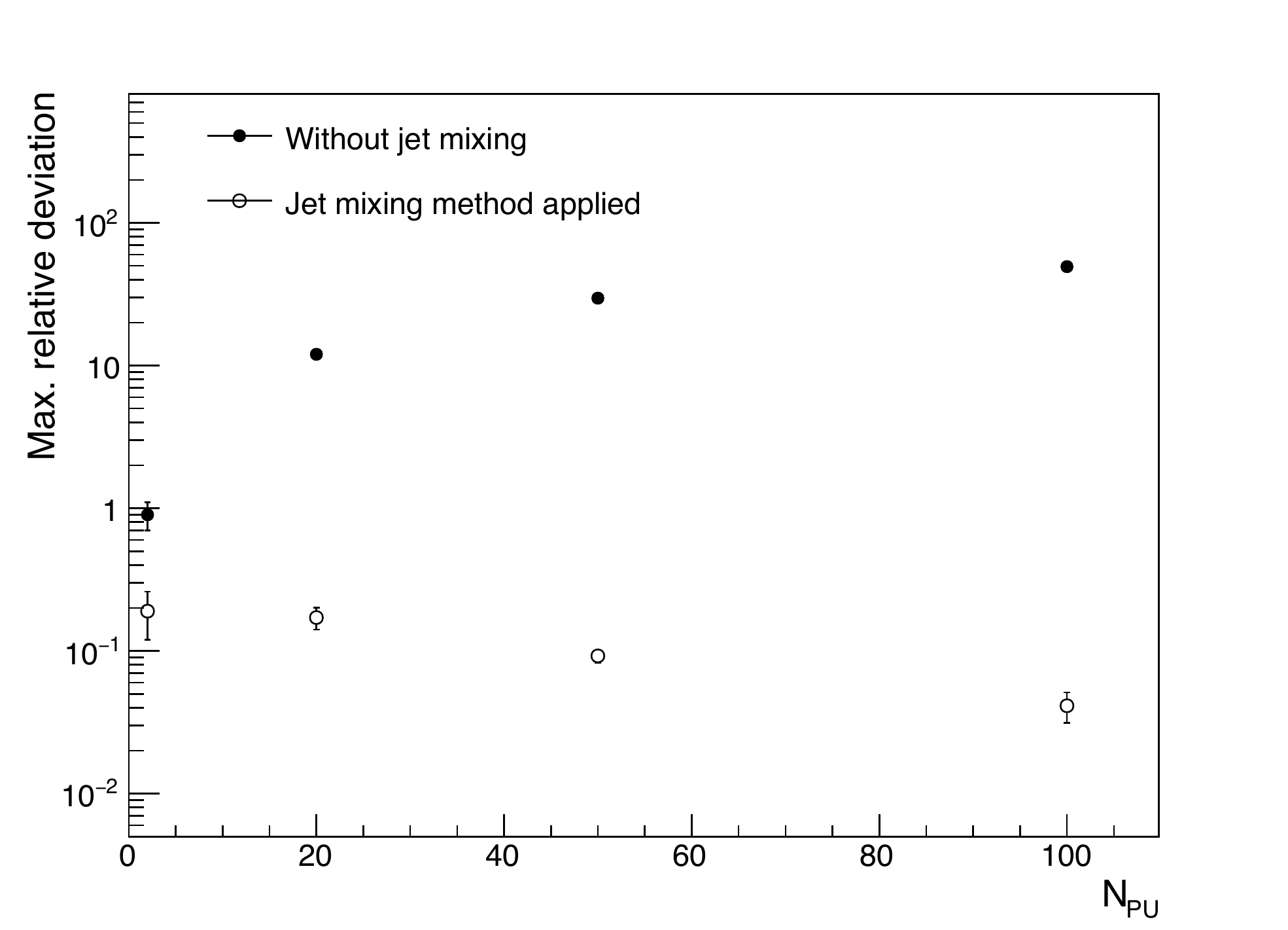}
  \caption{\it 
Maximal value of the 
relative deviation between  the pile-up corrected signal  and the true signal,  
with and without jet mixing, as a function of the number of pile-up collisions. Black dots: SoftKiller 
corrected result; open circles: jet  mixing method applied. }
\label{fig:plp_fig5}
\end{figure} 
The dashed blue curve is the 
high pile-up, 
SoftKiller-corrected result ($N_{\rm{PU}} = 50$ SK and $N_{\rm{PU}} = 100$ SK). 
As seen already in 
Fig.~\ref{fig:plp_fig2}b, this is 
far from the solid black curve in 
the lower-$p_T$ part of the 
spectrum. 
We regard the 
dashed blue curve as 
pseudodata in high pile-up. 
The long-dashed red curve is the 
jet mixing curve, obtained 
as described above by 
mixing the signal with the minimum 
bias sample.  The result of the mixing 
method is then given as the solid red curve  by a simple ``unfolding", 
defined by multiplying the 
signal by  the ratio  of the 
pile-up (dashed blue) curve to the 
mixing (long-dashed red) curve. 
We see that without appealing to any 
Monte Carlo method the true signal 
is extracted nearly perfectly 
from the mixed sample.  

In addition to the closure test 
 carried out above, we have 
checked the model dependence by 
applying the mixing procedure to  
different starting distributions, and 
verified that in this case as well the unfolding returns the true signal. 

In Fig.~\ref{fig:plp_fig5} we plot the maximal value of the 
relative deviation between 
 the pile-up corrected signal and  the 
 true signal ((corrected - true)/true), for 
the SoftKiller case 
 without jet  mixing (black dots) 
and for the case with the  jet  
mixing method applied (open circles), 
as  
a function of the number of pile-up collisions  $N_{\rm{PU}}$.   
We see that, while in the 
SoftKiller  case, in which  
the jet pedestal is removed,  
  the deviation from the 
true signal becomes larger  as $N_{\rm{PU}}$  
increases, the deviation 
does not increase with 
$N_{\rm{PU}}$  
 once the jet mixing method is applied to take account of 
the hard jets from pile-up.

\begin{figure}[htb]
\includegraphics[scale=0.4]{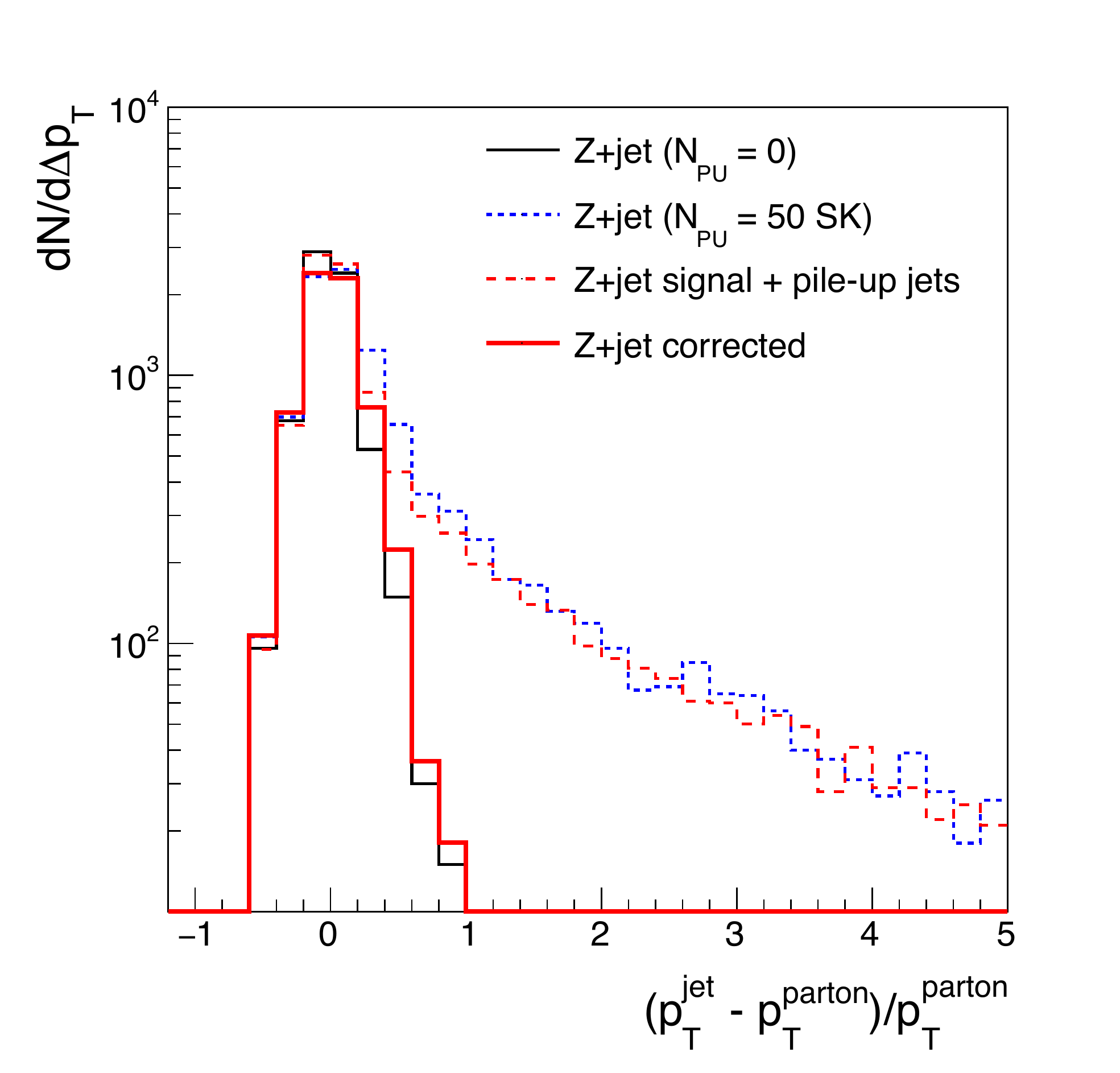}
\includegraphics[scale=0.4]{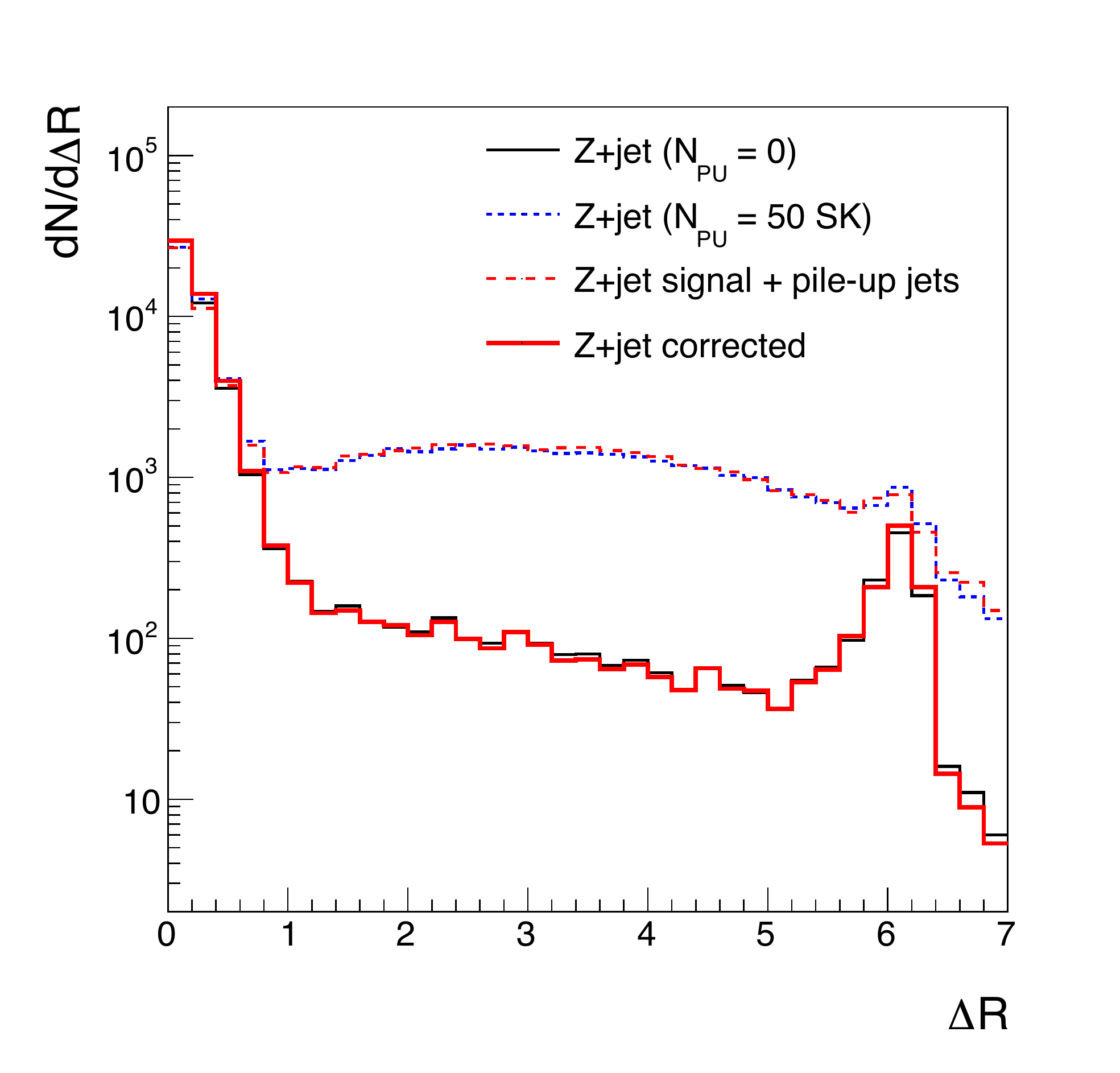}
  \caption{\it  Effects on jet resolution. Left: (a) the 
parton-jet $p_T$ correlation; 
right: (b) the $\Delta R $ distribution.}
\label{fig:plp_fig4}
\end{figure}

The main advantages of this 
approach are  that it can be used with   data 
recorded in high pile-up, and 
it does not  depend on  Monte 
Carlo algorithms for pile-up 
correction. In addition, it  is 
interesting to perform control 
checks by examining results 
for the jet resolution which we 
obtain from the  jet  mixing 
method. These are shown in 
Fig.~\ref{fig:plp_fig4}. 
Fig.~\ref{fig:plp_fig4}a reports the 
parton-jet $p_T$ correlation, and 
Fig.~\ref{fig:plp_fig4}b the 
distribution in $\Delta R = 
\sqrt{\Delta \phi^2 + 
\Delta \eta^2} $, where 
$\Delta \phi$ and  
$\Delta \eta$ are respectively the 
separation in azimuth and rapidity. 
We see that the features 
of the ``true" 
signal are well reproduced.

{\em 4.~Conclusions}. 
Current methods to deal with pile-up at the LHC employ 
precise vertex and track reconstruction, in regions where 
these are available,  and in general rely on Monte Carlo 
simulations to 
 model  pile-up for data comparisons. The use 
of Monte Carlo  event  generators 
 brings in a significant model dependence 
particularly in regions where these are not well constrained by 
measurements. 

In this 
paper we have discussed a different, data-driven  approach  to 
the treatment of pile-up, which   
makes use of minimum bias or  jet 
samples recorded from data taken in high pile-up runs,   constructs   
 mixing methods to extract the signal process, 
and thus circumvents   the model dependence implied by the use 
of Monte Carlo generators. 

The methodology  is general, and  can be applied  in 
measurements to restore  correlations between 
final-state particles. In such measurements two important 
kinds of  pile-up effects are present, exemplified in the 
case of  $Z$-boson  plus jets which we have used for illustration,  
 the jet $p_T$  pedestal 
and the misidentification of high-$p_T$ 
jets from independent pile-up events. While several methods exist 
to correct for the first effect (as well as for the related effect of the clustering 
into jets of overlapping soft particles), the second effect is not treated 
at present. We have proposed 
a jet-mixing  method which treats this, 
and we have shown  that it  allows one to successfully extract 
the signal process from the mixed  sample to within few percent.

The methods discussed in this paper can be applied to 
 the  high pile-up regime 
and do not require special  runs at low pile-up. The data  
 samples needed for  jet mixing are recorded at the same time as 
the signal events. There is therefore no loss in luminosity. The advantages 
are that one can access  the proper pile-up distribution and there is no 
need for pile-up reweighting. 

The use of these  methods thus implies good prospects 
both 
for precision Standard Model studies 
at moderate scales 
affected by pile-up, e.g.~in Drell-Yan and 
Higgs production~\cite{Cipriano:2013ooa,VanHaevermaet:2014ela,Langenegger:2015lra},  
and for searches for rare 
processes beyond Standard Model in high 
pile-up regimes. 

\vspace*{1cm} 
{\em Acknowledgments}. 
We thank M.~Cacciari, T.~Fruboes, E.~Gallo, 
P.~Gunnellini, S.~Haywood, B.~Murray, B.~Roland,   
  G.~Soyez  and P.~Van Mechelen 
for useful discussions. 
FH thanks the University of Hamburg and DESY for hospitality. 
The work of FH  is  supported in part  by  
  the DFG SFB 676 programme  Particles, Strings and the Early Universe. The work of HVH is 
funded by the 
Research Foundation - Flanders (FWO) in 
Belgium.

\end{document}